\documentstyle[11pt]{article}
\title{ {\bf Confining interquark potentials from nonabelian gauge theories coupled to dilaton}}

\author{Mohamed Chabab \thanks{correponding author:mchabab@ucam.ac.ma} ,  Latifa Sanhaji  \\ \\
{\it {\small LPHEA, Physics Department, Faculty of Science, Cadi Ayyad
University}}\\
{\it{\small  P.O. Box 2390, Marrakesh 40 000, Morocco}}}
\date{}
\begin{document}
\begin{titlepage}
\maketitle
\begin{abstract}
\baselineskip .7cm

Following a recently proposed confinement generating
mechanism, we provide a new string inspired model with a massive
dilaton and a new dilaton coupling function \cite{CSa}. By solving
analytically the equations of motion, a new class of confining
interquark potentials is derived which includes several popular
potential forms given in the literature.

keywords: Dilaton; confinement; quark potential.

\end{abstract}
\end{titlepage}
\textwidth 13.3cm
\parskip .5cm
\baselineskip .85cm
\textheight 21cm

\section{Overview}
Recently, the extension of gauge field theories by inclusion of dilatonic
degrees of freedom has evoked considerable interest. Particularly,
dilatonic Maxwell and Yang-Mills theories which, under some
assumptions, possess stable and finite energy solutions \cite{CT}.
Indeed, in theories with dilaton fields, the topological structure
of the vacuum is drastically changed compared to the non dilatonic
ones. It is therefore of great interest to investigate the vacuum
solutions induced by r-dependent dilaton field, through a string
inspired effective theory which may reproduce the main feature
of strong interactions: quark confinement. Recall that the dilaton
is an hypothetical scalar particle appearing in the spectrum of
string theory and Kaluza-Klein type theories \cite{GSW}. Along
with its pseudo scalar companion, the axion, they are the basis of
the discovery F-theory compactification \cite{Va} and of the
derivation of type IIB self duality \cite{Sen}. The main features
of a dilaton field is its coupling to the gauge fields through the
Maxwell and Yang-Mills kinetic term. In particular, in string
theory, the dilaton field determines the strength of the gauge
coupling at tree level of the effective action. In this context,
Dick \cite{Di} observed that a superstring inspired coupling of a
massive dilaton to the 4d $SU(N_c)$ gauge fields provides a
phenomenologically interesting interquark potential $V(r)$ with
both the Coulomb and confining phases. The derivation performed in
\cite{Di} is phenomenologically attractive since it provides a new
confinement generating mechanism. In this context, a general
formula of a quark-antiquark potential, which is directly related
to the dilaton-gluon coupling function, has been obtained in
\cite{Ch1}. The importance of this formula is manifest since it
generalizes the Coulomb and Dick potentials, and it may be
confronted to known descriptions of the confinement, particularly,
those describing the complex structure of the vacuum in terms of
quarks and gluons condensates. 

In this work, we shall propose a new effective coupling of a
massive dilaton to chromoelectric and chromomagnetic fields
subject to the requirement that the Coulomb problem still admits
an analytic solution. Our main interest concerns the derivation
of a new family of confining interquark potentials. As a by
product, it is shown that several popular phenomenological
potentials may  emerge from these low energy effective theories.

\section{The model}
Let us consider an effective field theory defined by the general
Lagrangian density:
\begin{equation}
{\cal L}({\phi},A)=
-\frac{1}{4F({\phi})}{G_{{\mu}{\nu}}^a}{G^{{\mu}{\nu}}_a}
+\frac{1}{2}\partial_\mu \phi \partial^\mu \phi  -V(\phi) +J_a^\mu
A_\mu^a
\end{equation}
 where $V(\phi)$ denotes the non perturbative scalar potential
of $\phi$ and $G^{\mu \nu}$ is the field strength in the
language of 4d gauge theory.\\
$F(\phi)$ is the coupling function depending on the dilaton
field. Several forms of $F(\phi)$ appeared in different
theoretical frameworks: $F(\phi)=e^{-k\frac{\phi}{f}}$ as in
string theory and Kaluza-Klein theories \cite{GSW};
$F(\phi)=\frac{\phi}{f}$ in the Cornwall-Soni model
parameterizing the glueball-gluon coupling \cite{CS,DF}. As to
Dick model, $F(\phi)$ is given by $F(\phi)=k+\frac{f^2}{\phi^2}$.
The constant f is a characteristic scale of the strength of the
dilaton/glueball-gluon. By using the formal analogy between the
Dick problem and the Eguchi-Hansen one \cite{GIOT}, we noted in
\cite{Ch1} that $f$ is similar to the $4d N=2$ Fayet-Illioupoulos
coupling in the Eguchi-Hansen model. It may be interpreted as the
breaking scale of the $U(1)$ symmetry rotating the dilaton field.

To analyze the problem of the Coulomb gauge theory augmented
with dilatonic degrees of freedom in (1), we proceed as follows:
first, we consider a point like static Coulomb source which is
defined in the rest frame by the current:
\begin{eqnarray}
J_a^\mu =g \delta (r) C_a \nu_0^\mu =\rho_a \eta_0^\mu
\end{eqnarray}

The equations of motion, inherited from the model (1) and emerging
from the static configuration (2) are given by:
 \begin{eqnarray}
 \left[ D_\mu , F^{-1} (\phi ) G^{\mu\nu}\right] = J^\nu
\end{eqnarray}
 and
\begin{eqnarray}
 \partial_\mu \partial^\mu \phi = -\frac{\partial
 V(\phi)}{\partial\phi}-\frac{1}{4} \frac{\partial F^{-1}(\phi)}{\partial
 \phi}G_a^{\mu\nu}G_a^{\nu\mu}
 \end{eqnarray}

By setting $G_a^{0i} = E^i \chi_a =-\nabla^i \Phi_a$, we derive,  after
some straightforward algebra, the important formula 
\cite{Ch1, Ch2},
 \begin{eqnarray}
\Phi_a (r) = \frac{-g C_a}{4\pi}\int dr \frac{F(\phi(r))}{r^2}
\end{eqnarray}
which shows that  confinement appears if the following
condition is satisfied:
\begin{eqnarray}
\lim_{r\to \infty} r F^{-1}(\phi(r)) = finite
\end{eqnarray}
Thereby the interquark potential reads as,
\begin{eqnarray}
U(r)&=&2\widetilde{\alpha}_s \int \frac{F(\phi(r))}{r^2} dr
\end{eqnarray}
with $\widetilde{\alpha} =\frac{g^2}{32\pi^2} \left( \frac{N_c -1
}{2N_c}\right)$

At this stage, note that the effective charge is defined by,
$$Q^a_{eff}(r)=\left(g\frac{C_a}{4\pi}\right) F(\phi(r))$$ thus the
chromo-electric field  takes the usual standard form: $$E_a
=\frac{Q^a_{eff}(r)}{r^2}$$Therefore, it is the running of the
effective charge that makes the potential stronger than the
Coulomb potential. Indeed if the effective charge did not run, we
recover the Coulomb spectrum.

To solve the equations of motion (3) and (4), we need to fix two
of the four unknown quantities $\phi(r)$, $F(\phi)$, $V(\phi)$
and $\Phi_a(r)$ in our model. We set $V(\phi)$ to $V(\phi)
=\frac{1}{2} m^2 \phi$ and introduce a new coupling function:
$$F(\phi)= \left(1-\beta\frac{\phi^2}{f^2}\right)^{-n}$$

By replacing $F(\phi)$ in equation (4), it becomes very difficult
to solve analytically. However since we are usually interested by
the large distance behavior of the dilaton field and its impact on
the Coulomb problem, an analytical solution in the
asymptotic regime is very satisfactory. Indeed, it is easily shown
that the following function:
\begin{eqnarray}
\phi =\left[\frac{f^2}{\beta}
-\left(\frac{\beta}{f^2}\right)^\frac{-n}{n+1}
\left(\frac{2n\alpha_s}{m^2}\right)^\frac{1}{n+1}
\left(\frac{1}{r}\right)^\frac{4}{n+1}\right]^\frac{1}{2}
\end{eqnarray}
solves (4) at large $r$. Therefore, thanks to the master formula
(5), we derive the potentials,
\begin{eqnarray}
\Phi_a (r)=-\frac{g C_a}{4\pi}\left( \frac{2n\beta\alpha_s}{m^2
f^2}\right)^\frac{-4n}{n+1} \frac{n+1}{3n-1} r^{\left(
\frac{3n-1}{n+1} \right)}
\end{eqnarray}
By imposing the condition (6), we obtain a family of confining
interquark potentials if $n\geq\frac{1}{3}$. If moreover, we
invoke the criterion of Seiler \cite{Se}, then the values of $n$
are constrained to the range $n\leq 1$. Therefore the confinement
in our model (1) appears for the coupling function
$\frac{1}{F(\Phi)}$ with $n \in \left[\frac{1}{3},1\right]$. Such
class of confining potentials is very attractive. Indeed, by
selecting specific values of $n$, we may reproduce several popular
interquark potentials: Indeed if $n=1$, we recover the confining
linear term of Cornell potential \cite{Ei}. Martin's potential
$(V(r)\sim r^{0.1} )$\cite{Ma} corresponds to $n=\frac{11}{29}$,
while Song-Lin interquark potential \cite{SL} and Motyka-Zalewski
potential \cite{MZ}, with a long range behavior scaling as
$\sqrt{r}$, are obtained by setting $n$ to $\frac{3}{5}$. Turin
potential \cite{Li} is also recovered for $n=\frac{5}{9}$. We see
then, that these phenomenological potentials, which gained
credibility only through their confrontation to the hadron
spectrum, are now supplied with a theoretical basis since they can
be derived from a low energy effective theory.

\section{Conclusion}
we have derived a family of electric solutions
corresponding to a string inspired effective gauge theory with a
 $r$-dependent massive dilaton and a new coupling function
$F(\phi)= \left(1-\beta\frac{\phi^2}{f^2}\right)^{-n}$. By
constraining the values of n via the Seiler criterion and the
condition of Eq.(6) we have shown the existence of a class of
confining interquark potentials. Also, several popular quark
potentials, which are successful in describing meson and baryon spectra,
 may emerge from such low energy effective theory.

\section*{Acknowledgements}

This work is supported by the program PROSTARS III, under the contract No. D16/04.

\end{document}